\documentclass[prl,superscriptaddress,preprint]{revtex4}%
\usepackage{amsmath}
\usepackage{amsfonts}
\usepackage{amssymb}
\usepackage{graphicx}%
\begin{document}

\title{Highly-flexible wide angle of incidence terahertz metamaterial absorber
}
\date{\today}\received{}
\author{Hu Tao}
\affiliation{Boston University, Department of Manufacturing
Engineering, 15 Saint Mary's Street, Brookline, Massachusetts,
02446}
\author{C. M. Bingham}
\affiliation{Boston College, Department of Physics, 140 Commonwealth
Ave., Chestnut Hill, MA 02467}
\author{A.~C.~Strikwerda}
\affiliation{Boston University, Department of Physics, 590
Commonwealth Ave, Boston, Massachusetts, 02215}
\author{D. Pilon}
\affiliation{Boston University, Department of Physics, 590
Commonwealth Ave, Boston, Massachusetts, 02215}
\author{D. Shrekenhamer}
\affiliation{Boston College, Department of Physics, 140 Commonwealth
Ave., Chestnut Hill, MA 02467}
\author{N. I. Landy}
\affiliation{Boston College, Department of Physics, 140 Commonwealth
Ave., Chestnut Hill, MA 02467}
\author{K. Fan}
\affiliation{Boston University, Department of Manufacturing
Engineering, 15 Saint Mary's Street, Brookline, Massachusetts,
02446}
\author{X. Zhang}
\email{xinz@bu.edu} \affiliation{Boston University, Department of Manufacturing
Engineering, 15 Saint Mary's Street, Brookline, Massachusetts,
02446}
\author{W. J. Padilla}
\affiliation{Boston College, Department of Physics, 140 Commonwealth
Ave., Chestnut Hill, MA 02467}
\author{R.~D.~Averitt}
\email{raveritt@physics.bu.edu} \affiliation{Boston University,
Department of Physics, 590 Commonwealth Ave, Boston, Massachusetts,
02215}

\begin{abstract}
We present the design, fabrication, and characterization of a
metamaterial absorber which is resonant at terahertz frequencies. We
experimentally demonstrate an absorptivity of 0.97 at 1.6 terahertz.
Importantly, this free-standing absorber is only 16 microns thick
resulting in a highly flexible material that, further, operates over
a wide range of angles of incidence for both transverse electric and
transverse magnetic radiation.
\end{abstract}

\maketitle

The initial impetus driving metamaterials research was the
realization that a negative refractive index n$(\omega)$ =
$\sqrt{\epsilon(\omega)\mu(\omega)}$ could be obtained by creating
subwavelength composites where the effective permittivity
$\epsilon(\omega)$ and effective permeability $\mu(\omega)$ are
independently specified \cite{smith1,shelby,veselago}. Additionally, metamaterials allow
for tailoring the impedance Z$(\omega)$ =
$\sqrt{\mu(\omega)/\epsilon(\omega)}$ in a manner not easily
achieved with naturally occurring materials. This newfound approach
to engineering n$(\omega)$ and Z$(\omega)$ offers unprecedented
opportunities to realize novel electromagnetic responses from the
microwave through the visible. This includes cloaks, concentrators,
modulators, spoof plasmons, with many more examples certain to be
discovered in the coming years \cite{schurig06a,pendry06,rahm2008,padilla06,chen06,garcia-vidal05}.

Quite recently, there has been considerable interest in creating
resonant metamaterial absorbers which, through judicious design of
n$(\omega)$ and Z$(\omega)$, offer the potential for near unity
absorption \cite{landy07,tao08,landy2008}. The idea is to minimize
the transmission and to simultaneously minimize, through impedance
matching, the reflectivity. This has been experimentally
demonstrated at microwave and terahertz frequencies
\cite{landy07,tao08,landy2008}. Recently, other approaches have been
theoretically put forward to extend these ideas to higher
frequencies or to increase range of angles of incidence over which
the absorptivity remains sufficiently large for applications
\cite{diem2008,puscasu2008,avitzour2008}.

While the idea of designing a resonant absorber could be of
potential use throughout the electromagnetic spectrum, this concept
is expected to be especially fruitful at terahertz frequencies where
it is difficult to find strong absorbers. Such absorbers would
clearly be of use for thermal detectors or as a coating material to
mitigate spurious reflections using continuous wave sources such as
quantum cascade lasers \cite{landy07,tao08,williams06,tonouchi07}. Progress has been promising
where the initial design yielded an absorptivity of 0.70 at 1.3 THz
\cite{tao08}. This work has been extended to a polarization insensitive
design with a demonstrated absorptivity of 0.65 at 1.15 THz \cite{landy2008}.

\begin{figure} [ptb]
\begin{center}
\includegraphics[width=2.8in,keepaspectratio=true]%
{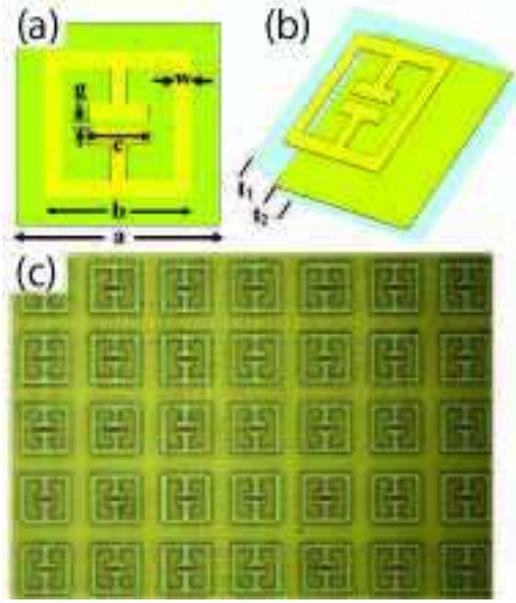}%
\caption{ THz metamaterial absorber consisting of two metallic
layers and two dielectric layers. (a) Electric SRR: unit cell a:
36$\mu$m, SRR side length b: 25.9$\mu$m, capacitor length c: 10.8
$\mu$m, capacitor gap g: 1.4$\mu$m, line width w: 3$\mu$m. (b)
Perspective view of the absorber. Each dielectric layer t$_{1}$ and
t$_{2}$ is 8$\mu$m thick.
(c) Photograph of a portion of the experimentally realized absorber.}%
\label{fig1}%
\end{center}
\end{figure}

In this letter, we experimentally demonstrate a resonant
metamaterial with an absorptivity of 0.97 at 1.6 THz. In
comparison to previous designs \cite{tao08,landy2008}, the current design has
several important advantages. Most importantly, the present design
is on a freestanding highly-flexible polyimide substrate 8$\mu$m
thick which enables its use in nonplanar applications as it can easily be wrapped around
objects as small a 6 mm in diameter. In addition, we demonstrate,
through simulation and experiment, that this metamaterial absorber
operates over a very wide range of angles of incidence for
transverse electric (TE) and transverse magnetic (TM)
configurations. Finally, the bottom layer of the absorber consists
of a continuous metal film which simplifies the fabrication in that,
for this two layer structure, precise alignment between the layers
is not required. We also discuss the relative importance of losses
in the metal and dielectric spacer layer.

Maximizing the absorption A is equivalent to minimizing both the
transmission T and reflectivity R in that A $=$ 1 - T - R. As has
been demonstrated \cite{landy2008}, in the limit that impedance matching to
free space is achieved (i.e. Z = Z$_{o}$ resulting in R $=$ 0), the
transmission reduces to T $=$ exp(-2n$_{2}$dk)= exp(-$\alpha$d)
where k is the free space wave vector, d is the sample thickness,
n$_{2}$ is the imaginary part of the  refractive index, and $\alpha$
is the absorption coefficient. Thus, impedance matching is a crucial
step yielding a transmission that is determined solely by the losses
in the slab of thickness d. In the case of a metamaterial absorber
the effective n$_{2}$ is determined by $\epsilon(\omega)$ and
$\mu(\omega)$. Thus, the design of a near-unity resonant
metamaterial absorber, $\epsilon(\omega)$ and $\mu(\omega)$ must be
optimized such that, at the desired center frequency, Z = Z$_{o}$
with n$_{2}$ as large as possible.

A compact metamaterial absorber consists of two metallic layers
separated by a dielectric spacer. The top layer consists of
an array of split ring resonators which is primarily responsible for
determining $\epsilon(\omega)$ while the bottom metallic layer is
designed such that the incident magnetic field drives a circulating
currents between the two layers. However, given the strong coupling
between the two layers, fine tuning of the geometry is required to
obtain the conditions described in the previous paragraph.
Fortunately, using full-wave electromagnetic simulation, rapid
convergence to a near optimal design is readily achieved.

Figure 1 presents such an optimized design which we have
subsequently fabricated and tested. The top layer (Fig. 1(a))
consists of an array of 200 nm thick Au electrically resonant split
ring resonators \cite{schurigAPL,padilla07}, with the dimensions as listed in the figure
caption. In the absence of a second metallic layer, this structure
yields a pure $\epsilon(\omega)$ response, and can be thought of as
an LC circuit as described elsewhere \cite{schurigAPL,padilla07}. A dielectric spacer layer
8$\mu$m thick separates the top and bottom metallic layers. The
bottom metallic layer is a continuous 200 nm thick Au film. As
Figure 1(b) shows, there is a second 8$\mu$m thick dielectric layer
which provides mechanical support but, being behind the continuous
Au film, does not contribute to the electromagnetic response. Figure
1(c) shows a photograph of a portion of the structure we have fabricated and
tested as detailed below.

The optimized structure presented in Figure 1 was obtained through
computer simulations using the commercial program CST Microwave
Studio$^{TM}$ 2006B.04. The frequency domain solver was utilized
where the Au portions of the metamaterial absorber was modeled as
lossy gold with a frequency independent conductivity $\sigma$ =
4.09$\times$10$^{7}$ S/cm. The 8$\mu$m thick dielectric layer was
modeled using the experimentally measured value of polyimide as this
is what is used in the subsequent fabrication. Specifically, a
frequency independent $\epsilon$ $=$ $\epsilon_{1}$ +
i$\epsilon_{2}$ $=$ 2.88 +i0.09 was used which corresponds to a loss
tangent tan($\delta$)$=$ $\epsilon_{2}$/$\epsilon_{1}$ $=$ 0.0313
\cite{tao08b}.  The amplitude of the transmission S$_{21}$ and
reflection S$_{11}$ were obtained and the absorption was calculated
using A = 1-R-T = 1-S$_{11}^{2}$-S$_{21}^{2}$ where, as expected for
the present design, S$_{21}$ is zero across the entire frequency
range due to the ground plane. The optimized structure presented in
Fig. 1 was obtained (simulating radiation at normal incidence)
through parameter sweeps of the dimensions of the SRR and the
dielectric spacer thickness. The optimized parameters are those
which yielded the lowest reflectivity at the design frequency of 1.6
THz.

\begin{figure}[ptb]
\begin{center}
\includegraphics[width=2.8in,keepaspectratio=true]%
{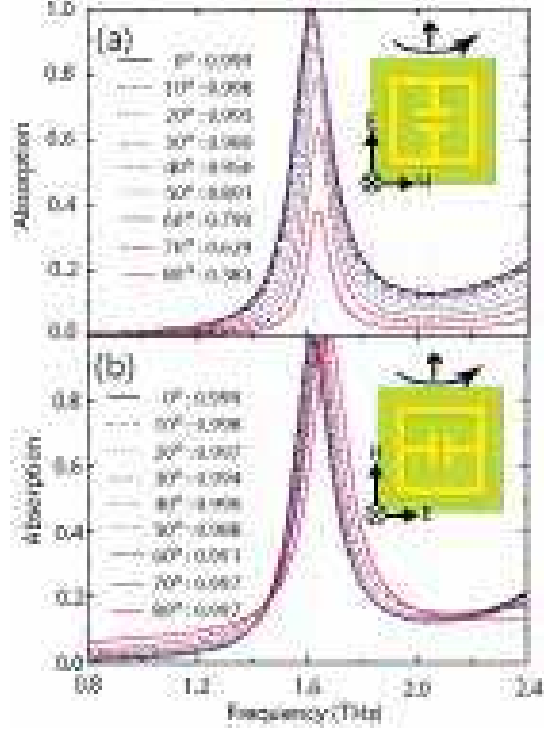}%
\caption{Simulations of the metamaterial absorber showing the
absorptivity as a function of frequency at various angles of
incidence for (a) TE  and (b) TM incident radiation. The insets
depict the orientation of the fields with respect to the SRR.
The labels for the curves show the angle of incidence and the corresponding peak absorptivity.}%
\label{fig2}%
\end{center}
\end{figure}

The simulated absorption as a function of frequency for the
optimized structure (Fig. 1) is presented in Figure 2 for TE (Fig.
2(a)) and TM (Fig. 2(b)) radiation at various angles of incidence.
For the TE case, at normal incidence a peak absorption of 0.999 is
obtained. With increasing angle of incidence, the absorption remains
quite large and is at 0.89 at 50$^{o}$. Beyond this there is a
monotonic decrease in the absorption as the incident magnetic field
can no longer efficiently drive circulating currents between the
two metallic layers. There is also a slight frequency shift of
$\sim$30 GHz from 0$^{o}$ to 80$^{o}$. For the case of TM radiation
shown in Fig. 2(b), the absorption at normal incidence is 0.999 at
normal incidence and remains greater than 0.99 for all angles of
incidence. In this case, the magnetic field can efficiently drive
the circulating currents at all angles of incidence which is
important to maintain impedance matching. The frequency
shift for TM radiation is $\sim$80 GHz from
0$^{o}$ to 80$^{o}$. As these simulations reveal, this MM absorber
operates quite well for both TE and TM radiation over a large range
of angles of incidence.

An additional aspect to consider in the design of metamaterial
absorbers are losses in the constituent materials comprising the
structure. As discussed in the introduction, one of the design
criteria is to obtain a large value of the imaginary part of the
effective refractive index. This necessitates having some losses in
the metal. Losses in the dielectric spacer are expected to
contribute as well. For example, in the limit of a perfect electric
conductor and a lossless dielectric, the absorption in the composite
in Fig. 1 is zero. However, losses in gold are sufficient to yield a
strong narrow band resonance as shown in Figure 2.

Fig. 3(a) and (b) show the calculated surface current density for a
TE wave at resonance. The currents are in opposite directions on SRR
and the ground plane as expected for a magnetic resonance. Figure
3(c) shows the absorption as a function of frequency for the design
in Fig. 1. The black curve assumes a lossless dielectric - i.e.
tan($\delta$) = 0. In this case, the peak absorption is 0.88 which
is smaller than the calculations in Figure 2. This suggests losses
in the dielectric contribute to increasing the absorption. For
example, increasing tan($\delta$) to 0.04 (blue curve, Fig. 3(c))
increases the absorption to 0.99 which is an increase of 0.11 in
comparison to a lossless dielectric. However, a point of diminishing
return is reached for larger values of tan($\delta$) (see Fig. 3(c))
in that the absorption decreases and the resonance broadens. These
results suggest that optimization of tan($\delta$) of the dielectric
spacer can maximize the metamaterial absorption. Further, it appears
the losses in polyimide (tan($\delta$) $=$ 0.0313) should contribute
$\sim$ 0.1 to the absorption of our metamaterial absorber as is
evident by comparing the black curves in Fig. 2(a) and 3(c).

\begin{figure}[ptb]
\begin{center}
\includegraphics[width=2.8in,keepaspectratio=true]%
{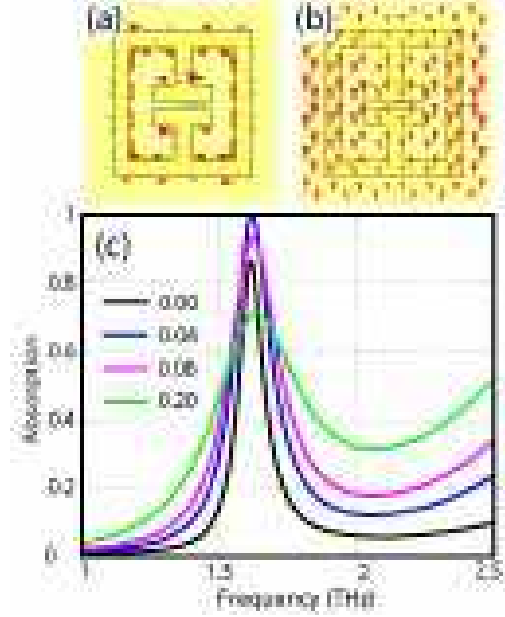}%
\caption{ (a) and (b) show the simulated on-resonance surface current density on the SRR
and ground plane, respectively, for TE incident radiation. The
currents are oppositely directed as expected for a magnetic
resonance. (c) Simulation of the absorption as a function of
frequency for various values of the dielectric loss tangent. The
curves are labeled with the value
of tan($\delta$) used in the simulation.}%
\label{fig3}%
\end{center}
\end{figure}

The free standing absorber structures were fabricated with a surface
micromachining process on flexible polyimide substrate using a
silicon wafer as the supporting substrate during the fabrication
process. Liquid polyimide (PI-5878G, HD MicroSystems$^{TM}$) was
spin-coated on a 2 inch silicon wafer to form the free standing
substrate. In this work, the polyimide was spin-coated at 1,700 rpm
and cured for five hours in an oven at 350°C in a nitrogen
environment yielding an 8$\mu$m thick polyimide layer. A 200 nm
thick Au/Ti film was e-beam evaporated on the polyimide substrate to
form the ground plane. Another 8$\mu$m thick polyimide layer was
spin coated on the top of the ground plane to form the polyimide
spacer and processed according to the procedure mentioned above. For
the SRR array, direct laser writing technology was chosen over
traditional mask contact lithography technology to improve the
patterning quality on the polyimide substrates. Shipley$^{TM}$ S1813
positive photoresist was first calibrated and then exposed with a
HeidelbergTM DWL 66 laser writer to pattern the top layer of
electric ring resonators. Another 200 nm thick Au/Ti film was e-beam
evaporated followed by rinsing in acetone for several minutes. The
metamaterial absorber fabricated on the polyimide substrate was
carefully peeled off of the silicon substrate at the end of
fabrication. Our samples show great mechanical flexibility and can
be easily wrapped around a cylinder with a radius of a few
millimeters.

\begin{figure}[ptb]
\begin{center}
\includegraphics[width=2.8in,keepaspectratio=true]%
{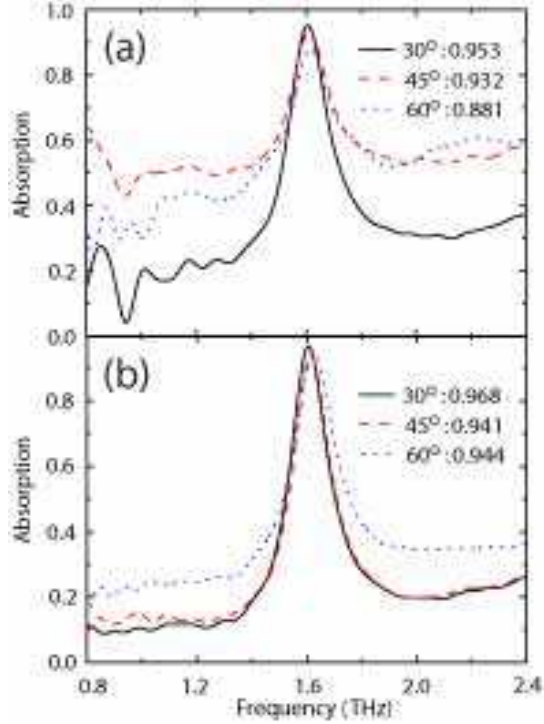}%
\caption{Experimentally measured absorption as a function of frequency for (a) TE and (b)
TM radiation at 30$^{o}$, 45$^{o}$, and 60$^{o}$ angles of incidence.}%
\label{fig4}%
\end{center}
\end{figure}

A Fourier transform infrared (FTIR) spectrometer was used to
experimentally verify the behavior of the absorber by measuring the
transmission and reflection over the frequency range of 0.6 THz to 3
THz with a resolution of 15 GHz. A liquid helium cooled bolometer
detector and 6 $\mu$m mylar beam splitter were used to optimize the
FTIR performance over the frequencies measured. Prior to
measurement, the free standing absorber samples were diced into 1 cm
$\times$ 1 cm squares. The aperture of the incident beam was 5 mm, which
is considerably smaller than the sample dimension. The sample was
mounted at normal incidence for the transmission measurement. As
expected, the transmitted intensity was essentially zero due to the
gold ground plane which blocks all radiation through the absorber.
The achievable incident angle for reflection measurements is
constrained within the range from 30° to 60° off-normal due to the
experimental limitations. The measurements were performed with
electric field perpendicular to the SRR gap to excite the electric
resonance. The absorption spectrum was easily obtained from the reflection
results (i.e. A $=$ 1-R).

The experimental results are displayed in Figure 4(a) and (b) for TE
and TM incident radiation, respectively. For the TE radiation, the
absorption peaks at 0.95 for an angle of incidence of 30$^{o}$
decreasing slightly to 0.88 at 60$^{o}$. This is in reasonable
agreement with the simulations though the experimental absorptivity
at 60$^{o}$ is $\sim$0.09 higher than for simulation. However, the
off-resonance absorptivity is quite large in disagreement
with the simulations. This may arise, in part, from scattering of
radiation from imperfections arising from the fabrication. For the
TM measurements the peak absorptivity is 0.968 at 30$^{o}$ angle of
incidence and only drops by 0.024 upon increasing to 60$^{o}$.
Further, the increase in the baseline absorption is much smaller in comparison to the TE measurements and
is in better agreement with simulations. A closer inspection of Fig. 4(b)
also reveals a slight increase in the resonance frequency with
increasing angle of incidence in agreement with simulation. Overall,
these results substantially confirm the simulation results demonstrating that
our MM absorber yields a large absorptivity over a broad range of
angles of incidence for both TE and TM radiation.

In summary, we have presented the design, fabrication, and
characterization of a highly flexible metamaterial absorber that,
experimentally, obtains an absorptivity of 0.96 at 1.6 THz, and
further, operates over wide angular range for TE and TM radiation.
Such a composite THz metamaterial may find numerous applications
ranging from the active element in a thermal detector to THz stealth
technology.

We acknowledge partial support from the Los Alamos National
Laboratory LDRD program, DOD/Army Research Laboratory
W911NF-06-2-0040, NSF EECS 0802036, and DARPA HR0011-08-1-0044.
The authors would also like to thank the
Photonics Center at Boston University for all of the technical
support throughout the course of this research.

\end{document}